\documentclass[aps,prd,10pt, tightenlines, twoside, secnumarabic, 
superscriptaddress,showpacs, preprintnumbers, showpacs, nofootinbib, 
twocolumn]{revtex4-1}

\pdfoutput=1
\usepackage{amsthm}
\usepackage{graphicx}
\usepackage{color}
\newcommand{\beq}{\begin{eqnarray}}
\newcommand{\eeq}{\end{eqnarray}}

\newcommand{\bmp}{\noindent\begin{minipage}{16cm}}
\newcommand{\emp}{\end{minipage}\vskip 7mm} 

\usepackage{dcolumn}
\usepackage{bm}
\usepackage{bbm}
\usepackage{subfigure}
\usepackage{slashed} 
\usepackage{mathrsfs}

\theoremstyle{definition}

\theoremstyle{plain}

\usepackage{epsfig}

\usepackage[ margin=5pt, font=normalsize,labelfont=bf,justification=raggedright]{caption}

\usepackage{hyperref}
\definecolor{rossoCP3}{cmyk}{0,.88,.77,.40}
\definecolor{verdeCP3}{rgb}{0.09765625, 0.57421875, 0.1015625}
\definecolor{bluCP3}{rgb}{0, 0.23, 0.67}
\hypersetup{colorlinks, bookmarksopen, bookmarksnumbered,citecolor=bluCP3, linkcolor=bluCP3, pdfstartview=FitH, urlcolor=bluCP3}
\def\lsim{\mathrel{\rlap{\lower4pt\hbox{\hskip1pt$\sim$}}
    \raise1pt\hbox{$<$}}}                
\def\gsim{\mathrel{\rlap{\lower4pt\hbox{\hskip1pt$\sim$}}
    \raise1pt\hbox{$>$}}}                

\newcommand{\bea}{\begin{eqnarray}}
\newcommand{\eea}{\end{eqnarray}}

\newcommand{\ba}{\begin{eqnarray}}
\newcommand{\ea}{\end{eqnarray}}

\newcommand{\be}{\begin{eqnarray}}
\newcommand{\ee}{\end{eqnarray}}

\begin{document}
\font\secret=cmr10 at 0.8pt
\title{ ~~\\  Daily Modulation as a Smoking Gun of Dark Matter with Significant Stopping}
%

\author{Chris Kouvaris}
\author{Ian M. Shoemaker}
\email{shoemaker@cp3-origins.net} 
\affiliation{CP$^{3}$-Origins \& Danish Institute for Advanced Study  DIAS, University of Southern Denmark, Campusvej 55, DK-5230 Odense M, Denmark}
\email{kouvaris@cp3.dias.sdu.dk} 

\begin{abstract}
We point out that for a range of parameters, the flux of DM may be stopped significantly by its interactions with the Earth. This can significantly degrade the sensitivity of direct detection experiments to DM candidates with large interactions with terrestrial nuclei. We find that a significant region of parameter space remains unconstrained for DM $\lesssim $ a few GeV. For DM candidates with moderate levels of stopping power, the flux of DM may be blocked from below but not above a detector thereby producing a novel daily modulation. This can be explored by low threshold detectors placed on the surface or in shallow sites in the south hemisphere.
\\[.1cm]
{\footnotesize  \it Preprint: CP3-Origins-2014-019 DNRF90,DIAS-2014-19}
 \end{abstract}

\maketitle

\section{Introduction}

At present many experimental searches use underground detectors in order to be able to detect dark matter particles, while being shielded from unwanted surface backgrounds. 
Most detection techniques are based on WIMP-nucleus collisions that produce sufficient recoil energy to be detected either as phonons, ionization or scintillation light. However it is clear that in order to detect a WIMP it is not merely sufficient to have large detector exposures, since one must also ensure that a WIMP has enough kinetic energy to produce a recoil energy above threshold. It is therefore crucial to know how much kinetic energy WIMP lose as they travel from the halo, through the Earth, and finally arrive at the detector. Obviously due to the higher density of the Earth (compared to outer space) most of the energy loss for a WIMP might take place along the distance traveling underground on the way to the detector. It is well known that Strongly Interacting Massive Particles lose enough energy to reach underground detectors with energies way below the required to trigger a signal~\cite{Fargion:2005ep,Khlopov:2006uv}.  
 
In this paper we survey the stopping force that WIMPs experience as they travel underground under a variety of assumptions for the form of the DM-nucleus interaction: electric/magnetic dipoles, light mediator exchange, and contact interactions. Previous work has studied the effect of nuclear stopping for contact interactions~\cite{Collar:1993ss,Hasenbalg:1997hs,Zaharijas:2004jv,Mack:2007xj}, and has been briefly touched upon for millicharged DM~\cite{Foot:2003iv,Cline:2012is} and electromagnetic dipole moments~\cite{Sigurdson:2004zp}. We find that although these type of particles might  have large enough cross sections to be easily detected in underground experiments like LUX, significant deceleration during their travel underground can invalidate this possibility simply because of their deceleration to energies below the threshold of detection. We study in detail electronic and nuclear stopping power. As we shall argue, there is significant parameter space for contact and light mediator WIMPs that will not be able to be excluded in the current underground detectors irrespective of the accumulated exposure. We show that low threshold detectors in shallow sites can actually probe this parameter space and we demonstrate that depending on the strength of the interaction a diurnal modulation of a dark matter signal could potentially be observed.
\section{Stopping Power}
It is not an easy task to estimate the stopping power of millicharged WIMPs. First of all, the well known Bethe formula is not applicable in this case. This is because WIMPs are moving too slowly to be able to ionize an atom as they move through matter underground. In fact the energy is so low that WIMPs can decelerate by i) electronic Coulomb interactions in insulators ii) electronic Coulomb interactions in metals iii) nuclear recoil. We find that the most efficient of all is nuclear stopping. Therefore the reader that is interested in just the results can skip the section about insulators (Sec.~\ref{sec:ins}) and conductor electronic stopping (Sec.~\ref{sec:cons}) and move to nuclear stopping (Sec.~\ref{sec:nuc}), which is the one that dominates the stopping process in our case. However for completeness we now examine each case in turn.
\subsection{Insulators}
\label{sec:ins}
Different experiments lie on different sites with different ground compositions. However by considering for example the site of the LUX experiment which is located at the Homestake mine, chemical element composition analysis~\cite{Mei:2009py} suggests the significant existence of insulating chemical compounds such as $SiO_2$, $Al_2O_3$, $FeO$, etc. Scattering of WIMPs with electrons from these chemical compounds can excite the molecules while the WIMPs lose energy. It is very difficult to estimate with high accuracy what the stopping power will be in this case since the exact oscillator strengths (in infrared frequencies) of all these compounds are needed.

Here we present an estimate of  the electronic stopping power of insulators for slowly moving projectiles based on Bohr's oscillator model~\cite{Sigmund}. In this case, if the projectile is slowly moving and it cannot ionize an atom, the stopping power can be approximated by
\begin{equation}
\frac{dE}{dx}=-\frac{4\pi \varepsilon^2 e^4}{m_e v^2}nZ \mathcal{L} ,
\end{equation}
where $m_e$ is the electron mass, $v$ the projectile velocity, $n$ the atomic number density, $Z$ the number of electrons per atom and $ \mathcal{L}$ is a dimensionless number tabulated in~\cite{Sigmund} as a function of the parameter $2m_ev^2/\hbar \omega$ where $\omega$ is the oscillator frequency. If one recalls that $E=1/2m_Xv^2$, the above equation is easily solved as
\begin{equation}
E_{in}^2-E_f^2=(4\pi)^3 \frac{m_X}{m_e} \varepsilon^2 \alpha^2 nZ\overline{\mathcal{L}}L,
\end{equation}
where $E_{in}$ is the initial energy of the millicharged particle, $E_f= E_{\rm thr}(m_X +m_N)^2/(4m_Nm_X)$ is the final energy that can lead to a maximum recoil of $E_{\rm thr}$ for a collision between the WIMP and a nucleus of mass $m_N$, $\alpha$ is the fine structure constant, $L$ the distance the particle has traveled underground,  and $\overline{\mathcal{L}}$ the average value of $ \mathcal{L}$. Let us consider for simplicity the case where matter is made of atomic  hydrogen, where $\omega \sim 10$ eV. In Fig.~1  we show the lowest value of $\varepsilon$, where $q=\varepsilon e$ ($e$ being the electron charge and $q$ the millicharge), above which particles decelerate after 1.6 km (LUX's depth) to energies that  produce always recoil below the 3 keV threshold of the experiment, upon assuming an average density of the Earth crust of $2.7 \text{g}/\text{cm}^3$, and a WIMP velocity of $\sim 300\text{km}/\text{s}$ (for illustration). Although in Fig.~1 it seems that electronic stopping of insulators can decelerate slightly more effectively millicharged particles compared to  nuclear stopping, in reality this never happens. The first reason is because matter above the detector is not in the form of pure hydrogen. Heavier elements and chemical compounds can have frequencies larger than 10 eV we used for hydrogen. This will  significantly reduce $\mathcal{L}$ with a concomitant reduction in the stopping power. In addition to this, there will also be millicharged particles in the velocity distribution with velocities below $\sim 300~\text{km}/\text{sec}$ for which the value of $\mathcal{L}$ will also be substantially smaller. Therefore although in Fig.~1 it appears that electronic insulator stopping could do slightly better in terms of deceleration compared to nuclear stopping, low velocity WIMPs do not stop effectively thus invalidating electronic stopping based on insulators as the dominant stopping mechanism.

 Experiments of antiproton projectiles on the insulator LiF have determined a stopping power with a peak of about $8~\text{eV}/\AA$ at an energy of $100~\text{keV}$ that drops to $\sim 1 \text{eV}/\AA$ for a velocity of antiproton $300\text{km}/\text{sec}$~\cite{Moller}. A similar experiment using protons instead of antiprotons gives a higher stopping power of $\sim 13~\text{eV}/\AA$ peaking again at $\sim 100\text{keV}$. The difference between the stopping power of protons and antiprotons is attributed to the fact that as they travel through matter protons can effectively ``share" and exchange electrons with atoms of the target leading to a higher stopping power. On the contrary this mechanism is absent in the case of antiprotons since they carry the same charge as the electrons and thus cannot form bound states. The case of millicharged particles is closer to antiprotons than protons simply because their tiny electric charge corresponds to large Bohr radii and we therefore expect that it is unlikely that millicharged particles can ``share" electrons with atoms as they move through matter. In order to find the actual stopping power of a millicharged particle we have to scale the stopping power of antiprotons by $\varepsilon^2$. By considering the stopping power of antiprotons with velocity $300\text{km}/\text{sec}$ scaled by $\varepsilon^2$, we show in Fig.~1 above what value of $\varepsilon$ stopping by a typical insulator like LiF can decelerate millicharged particles to energies that cannot recoil with energies above the 3 keV threshold of LUX. It is evident from Fig.~1 that electronic stopping from insulators cannot do better than nuclear stopping. One might be puzzled by the fact that in Fig.~1 the curve of LiF stopping is higher than the nuclear stopping since experimentally it is the total stopping that can be measured and therefore total (electronic and nuclear) stopping should always be larger than just one part of it i.e. the nuclear one. However, one should keep in mind that the experiment has been performed on LiF while in Fig.~1 we consider oxygen nuclei  for nuclear stopping (which turns out giving the highest nuclear stopping above the LUX detector). Although it is impossible to check all possible compounds, it appears that the electronic stopping of insulators is not more effective than nuclear stopping which we study in detail in Sec.~\ref{sec:nuc}.

\subsection{Conductors}
\label{sec:cons}
The electronic stopping power might be in principle significantly different for slowly moving projectiles if they can scatter off conductors. Slowly moving particles that do not have sufficient energy to excite electrons face no problem like this in conductors. Free electrons in metals can contribute significantly to stopping since there is no energy gap and therefore even slow moving particles can recoil effectively by transmitting small amounts of energy that effectively add up to a sizeable stopping. Although metals do not often appear in pure form in rock formations, there can indeed be conducting media such as for example graphite. Before we review the graphite stopping power, we can actually estimate the generic stopping power of metalic material such as for example iron. The stopping power of slow moving projectiles inside metals has been studied within the framework of the free electron gas by~\cite{Lindhard}. The stopping power is given by
\begin{equation}
\frac{dE}{dx}=-\frac{4 \varepsilon^2 e^4 m_e^2}{3 \pi} v C_1(\chi), \label{f1}
\end{equation}
where 
\begin{equation}
C_1(\chi)=\int_0^1\frac{z^3dz}{(z^2+\chi^2f_1(0,z))^2}.\label{f2}
\end{equation}
The function is defined in~\cite{Lindhard} 
\begin{equation}
f_1(0,z)=\frac{1}{2}+\frac{1}{4z}(1-z^2) \ln \left | \frac{z+1}{z-1} \right |. \label{f3}
\end{equation}
$\chi$ is defined as $\chi^2=e^2/(\pi v_F)$, $v_F$ being the Fermi velocity of the free electrons in the conductor. We should emphasize that the above stopping power is valid so long
$v<<v_F$. The Fermi velocity can be easily estimated as $v_F=(3 \pi^2 n)^{1/3}/m_e$ where $n$ is the number density of free electrons that in the case of a typical metal like iron is  $n=2.7  \cdot 0.6 N_a/(56~\text{cm}^3)$. Iron  has approximately 0.6 free electrons per atom at room temperature, and 56 atomic number. $N_a$ is the Avogadro number and we assumed a crust density of $2.7~\text{g}/\text{cm}^3$. 
By using the fact that $v=\sqrt{2E/m_X}$, we can solve  Eq.~(\ref{f1}) getting
\begin{equation}
\sqrt{E_{\rm in}}-\sqrt{E_{\rm f}}=\frac{2\sqrt{2}\varepsilon^2e^4m_e^2}{3\pi\sqrt{m_X}}C_1(\chi)L,
\end{equation}
where $L=1.6$ km is a typical distance a particle can travel underground to reach the detector. Using iron as a typical metal and the above result we can estimate the value of $\epsilon$ above which a particle decelerates to kinetic energies that can produce recoil energies scattered off $^{132}$Xe less than e.g. 3 keV. This $\varepsilon$ value is plotted as a function of DM mass in Fig.~1. We remind the reader that as in Sec.~\ref{sec:ins}, here the WIMP velocity has been taken to be $300~\text{km}/\text{sec}$. As can be seen, electronic stopping is not more effective than nuclear one at this velocity. We found that for $v=700~\text{km}/\text{sec}$ which is the maximum velocity a WIMP can have, electronic stopping becomes slightly more effective than nuclear stopping. However, this is not a realistic possibility since the estimate that we present in Fig.~1 assumes that the entire 1.6 km above the detector is conducting iron. It is easy to show that even a large layer of conducting matter cannot provide an electronic stopping that is more efficient than the nuclear one.

A more realistic scenario is that of the graphite since it is often abundant in rock formation. Graphite is pure carbon that is conducting only along a plane. It is an insulator along the third dimension. Graphite's stopping power is expected to be significantly less than that of a pure metal. This is because the number of electrons contributed per atom is $\sim 10^{-4}$ and in addition graphite is not conductive in any direction but only along a plane. Our conclusion is that electronic stopping is in general less efficient at decelerating millicharged particles compared to nuclear stopping. Let us now proceed to a detailed examination of nuclear stopping of WIMPs.

\section{Nuclear Stopping}
\label{sec:nuc}

\subsection{Millicharged WIMPs and Light Mediators}
The nuclear stopping power of millicharged DM can be evaluated as 
\be 
\frac{dE}{dx} = -n_{N} \int_{0}^{E_{R}^{max}} \frac{d\sigma}{dE_{R}} E_{R}dE_{R}, \label{m0}
\ee
where
\be \frac{d\sigma}{dE_{R}} = \frac{8 \pi \alpha_{EM}^{2}\varepsilon^{2} Z^{2}m_N}{v^{2}(2m_NE_{R}+\mu_0^2)^2}. \label{m02}
 \ee
The above cross section corresponds to the scattering of the millicharged particle via a screened Coulomb potential of the form $V=e_1e_2 e^{-\mu_0r}/r$, where $e_{1,2}$ are the charges of the  WIMP and the nucleus and $\mu_0=1/a$ is the inverse of the characteristic Thomas-Fermi radius $a=0.8853 a_0/Z_2^{1/3}$ ($a_0$ being the Bohr radius) beyond which the nucleus charge is screened. Although Eq.~(\ref{m0}) can be integrated analytically using Eq.~(\ref{m02}), we found that it is an excellent approximation to consider Coulomb instead of Yukawa potential for the scattering, cutting the integral to a minimum recoil energy $E_{R}^{min}=1/(2m_Na_0^2)$. It is understood that at lower recoil energies the nuclear charge is screened. We choose to use this approximation since it will simplify the analysis in the case of the electric and magnetic dipole interactions. Using the aforementioned approximation, Eqs.~(\ref{m0}) and (\ref{m02}) become
\be \frac{dE}{dx} = -n_{N} \int_{E_{R}^{min}}^{E_{R}^{max}} \frac{d\sigma}{dE_{R}} E_{R}dE_{R}, \label{m1}
\ee
 where the cross section takes a familiar form for long-range scattering
 \be \frac{d\sigma}{dE_{R}} = \frac{2 \pi \alpha_{EM}^{2}\varepsilon^{2} Z^{2}}{m_{N}v^{2} E_{R}^{2}}. \label{m2}
 \ee
Combining Eqs.~(\ref{m1}) and (\ref{m2}) we get
\be 
\frac{dE}{dx}=-\frac{A}{E} \int_{E_{R}^{min}}^{E_{R}^{max}}\frac{1}{E_R}dE_R, \label{m3}
\ee
where $A=\pi \alpha_{EM}^{2}n_N \varepsilon^2Z^2 m_X/m_N$. In the above equation we have traded $v^2$ for $E$. $E_{R}^{min}$ is the minimum recoil energy for the scattering and it is determined by the fact that the charge of the nucleus cannot be seen if the impact parameter is larger than the size of the atom. Therefore $E_{R}^{min}=1/(2m_Na_0^2)$. $E_{R}^{max}=4m_Xm_NE/(m_X+m_N)^2$  is the maximum recoil energy given a WIMP of energy $E$. Upon integration Eq.~(\ref{m3}) becomes
\be
\frac{dE}{dx}=-\frac{A}{E}\ln (bE), \label{m4}
\ee
where $b \equiv \frac{8 m_{N}^{2} m_Xa_{0}^{2}}{\left(m_{N}+m_X\right)^{2}}$. Finally we can integrate Eq.~(\ref{m4}) to find
\be
 {\rm Ei}\left( 2 \ln b E_{in}\right)- {\rm Ei}\left( 2 \ln b E_{f}\right)=\frac{\pi n_N \alpha_{EM}^{2} \varepsilon^2 Z^2b^2 m_XL}{m_N}, \label{m5}
\ee
where $L$ is the distance the particle has traveled underground, $E_{in}$ and $E_f$ are the initial and  final energies of the particle  
 %
 %
 and the exponential integral function is defined as ${\rm Ei}(x) \equiv - \int_{-x}^{\infty}\frac{e^{-t}}{t}dt$.  Eq.~(\ref{m5}) can be trivially solved for $\varepsilon$, which is interpreted as the minimum value of $\varepsilon$ above which the millicharged particle will decelerate to energy $E_f$ within distance $L$. We found that of the relevant terrestrial elements, the contribution of terrestrial oxygen gives the most significant contribution to nuclear stopping. Using~\cite{Mei:2009py} we estimate a $\sim 48\%$ oxygen in matter above for example the LUX detector.

\begin{figure*}[t!]
  \centering
                 \includegraphics[width=0.55\textwidth]{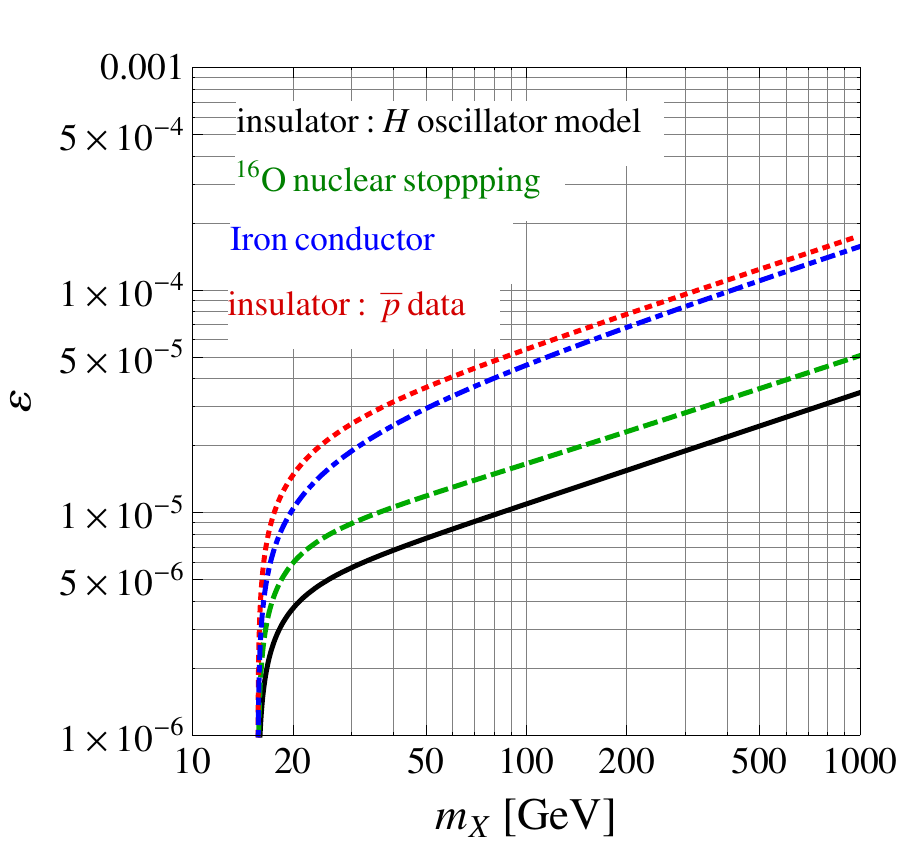}

   \caption{Here we compare different stopping processes i.e. oxygen nuclear stopping, electronic stopping on conducting iron, and electronic stopping based on insulators using a hydrogen oscillator model and  data from an antiproton experiment that uses LiF as a target. See the text for the detailed comparison of all different processes.    }
  \label{fig:nuc}
\end{figure*}

 Examining just the terrestrial oxygen contribution to the nuclear stopping, we find in Fig.~\ref{fig:nuc} that millicharged DM has significant stopping in matter. The inclusion of the other elements only moderately shifts the result to smaller values of $\varepsilon$.
 
In the case where WIMPs are not millicharged but interact with nuclei via exchange of light mediators (with finite rest mass), the cross section is given by Eq.~(\ref{m02}) where $\mu_0$ is now the larger between the mass of the mediator and the screening energy scale defined in Eq.~(\ref{m02}). As we will show in the section where we present our results, for mediators ranging from the eV scale to the MeV scale, the $\varepsilon$ above which WIMPs decelerate effectively is within a factor of 3 of the millicharged case (where the mediator has zero mass).

\subsection{Contact Interactions}
Contact interactions arise when the exchanged momentum is much smaller than the mass of the exchanged mediator. In this case the cross section is 
\be
\frac{d\sigma}{dE_R}=\frac{m_N \sigma_N}{2\mu_N^2 v^2}=\frac{m_N \sigma_n A^2}{2 \mu_p^2 v^2}, \label{contact}
\ee
where $\sigma_n$ and $\sigma_N$ correspond respectively to WIMP-nucleon and WIMP-nucleus cross sections, and $\mu_{p,N}$ refers to to the corresponding reduced masses between the WIMP and nucleon or nucleus. Using Eqs.~(\ref{m1}) and (\ref{contact}), we find that
\be
\ln\frac{E_{in}}{E_f}=\frac{2n_N \sigma_p A^2 \mu_N^4 L}{m_Xm_N \mu_p^2},
\ee
where once again $L$ is the length traveled underground, $E_{in}$ the initial energy of the WIMP and $E_f$ the final energy that leads to recoil less than $E_{\rm thr}$.

 \subsection{Electric Dipoles}

Many DM models lead to WIMPs acquiring an electric dipole moment, $\mathscr{L}_{EDM} = \frac{1}{2} d_{\chi} \overline{X} \sigma_{\mu \nu} \gamma^{5} X F^{\mu\nu}$. Electric dipoles have a cross section~\cite{Barger:2010gv}
\be
\frac{d\sigma_E}{dE_R}=\frac{A}{EE_R}, \label{cross_electric}
\ee
where $A=\frac{m_X}{8 \pi} d_{\chi}^2 Z^2 e^2\frac{S+1}{3S}|G_E(q^2)|^2$, $d_{\chi}$ being the DM electric dipole moment, $S$ being the spin of the WIMP and $|G_E(q^2)|$ is a nuclear form factor defined in~\cite{Barger:2010gv}. Using the above equation and following the steps of Eqs.~(\ref{m1}), (\ref{m3}), we get the stopping power to be determined by
\be
E_{in}-E_f+\frac{E_R^{min}}{\gamma}\ln \left (\frac{ E_f}{ E_{in}} \right )=\gamma n_N A L, \label{stop_electric}
\ee
where $\gamma=\frac{4m_Xm_N}{(m_X+m_N)^2}$. Recall that $A$ depends on $d_{\chi}$ and therefore the above equation sets a minimum value of electric dipole above which particles decelerate below energy $E_f$ after the have traveled distance $L$ underground. We should emphasize here that we choose the value of $E_f$ that gives a recoil below the threshold of each underground experiment into consideration. The recoil energy is at most $E_R^{max}=\gamma E_f$ and therefore $E_f$ is determined by demanding the recoil energy to be smaller than the threshold of the detector $E_R^{max}<E_{\rm thr}$.
\subsection{Magnetic Dipoles}
DM with a magnetic dipole, $\mathscr{L}_{MDM} = \frac{1}{2} \mu_{\chi} \overline{X} \sigma_{\mu \nu}X F^{\mu\nu}$, can arise in models in which the DM is coupled to heavy charged particles (see e.g.~\cite{DelNobile:2012tx,Frandsen:2013bfa} for a model and recent phenomenology of MDM).  The cross section for nuclear scattering is~\cite{Barger:2010gv}
\be
\frac{d\sigma_M}{dE_R}=\frac{B}{E_R} \left (1-\beta\frac{E_R}{E} \right ) \label{cross_magnetic}
\ee
where $B=\frac{e^2\mu_{\chi}^2}{4\pi}\frac{(S+1)}{3S}Z^2|G_E(q^2)|^2$, $\beta= \frac{1}{2}+ \frac{1}{4} \frac{m_X}{m_N}-\frac{I+1}{6IZ^2}\frac{\mu^2}{\mu_0^2}\frac{G_M(q^2)^2}{G_E(q^2)^2}\frac{m_Nm_X}{m_p^2}$, and $\mu_{\chi}$ is the magnetic dipole moment of the WIMP. The last term in the definition of $\beta$ corresponds to the spin dependent part of the cross section and it is always subdominant to the spin independent part. $I$ and $\mu$ denote respectively the spin and the magnetic moment of the nucleus, $m_p$ is  the proton mass, $\mu_0=e/(2m_p)$ is the nuclear magneton  and $G_M$ is a magnetic form factor that to first approximation can be taken equal to $G_E$. Using the above equation and following the steps of Eqs.~(\ref{m1}), (\ref{m3}), we get that the stopping power is determined by
\be
\int \frac{2EdE}{2 \gamma' E^2+ \beta E_R^{2min}-2E_R^{min}E}=-n_NB \int dx, \label{magn}
\ee
where $\gamma'=\gamma-\beta\gamma^2/2$ and $\gamma$ has been defined in the electric dipole subsection. Although Eq.~(\ref{magn}) has a analytic solution, it is easy to see that the dominator in the integral of the left hand side is dominated by the first term. Therefore the stopping equation takes the simple form
\be
\ln\frac{E_{in}}{E_f}\simeq n_NB\gamma'L. \label{stop_magnetic}
\ee
Eq.~(\ref{stop_magnetic}) can be used to set a minimum value on the magnetic moment of the WIMP $\mu_{\chi}$, above which the WIMP decelerates to a final energy $E_f$ that can give recoil energy below the threshold. As before, the relation $\gamma E_f=E_{\rm thr}$ determines the $E_f$ that gives maximum recoil up to $E_{\rm thr}$.
 
We should mention (as it can also be seen) that for the derivation of the formulas for nuclear stopping both for the electric and the magnetic dipole cases we take the corresponding form factors to be one. However we have found that including the exact form does not change significantly the stopping range.

\section{Results}

In the top panel of Fig.~2 we present the allowed/constrained phase space for millicharged and light mediator dark matter in the $(\varepsilon-m_X)$ plane. The constraints have been drawn taking into account the nuclear stopping of particles
once they are underground. We include LUX~\cite{Akerib:2013tjd}, DAMIC~\cite{Barreto:2011zu}, the CDMS shallow site~\cite{Abrams:2002nb}, CRESST-1~\cite{Altmann:2001ax}, SLAC accelerator searches~\cite{Prinz:1998ua,Davidson:2000hf}, the X-ray Quantum Calorimetry Experiment (XQC)~\cite{Erickcek:2007jv}, and the RRS balloon experiment~\cite{Rich:1987st}. Analysis and experimental details of the various experiments can be found in the Appendix. The figure also 
includes the strong constraints from the CMB (blue dashed line) based on the fact that values of $\varepsilon$ higher than the line lead to significant change of the acoustic peaks of the CMB spectrum (see e.g.~\cite{Dolgov:2013una,Dvorkin:2013cea}).
The constraints from underground experiments are given in the form of a band. Dark matter particles lying below the band are not excluded because they do not produce enough number of detectable events. Dark matter particles above the band however also evade direct detection constraints since here the nuclear stopping power is sufficient to decelerate the particles to energies below threshold.
It is amazing for example that for $\varepsilon \sim 10^{-4}$, even with a mass 
$m_X=100$ GeV, LUX would be unable to detect such a particle. Turning to the CDMS-1 data (situated at a depth of $\sim$10.8 m), we see that the improvement in constraining larger interactions is modest  (under the assumption that the shallow site background is well understood). However, in the case of millicharged dark matter, for WIMP masses roughly larger than 10 GeV, yet larger values of $\varepsilon$ are excluded by the RRS balloon borne experiment~\cite{Rich:1987st} (for which there is no appreciable stopping) and by the aforementioned 
CMB constraint (labeled by the $m_{\phi} =0$ curve for the millicharged DM case). However, one can see that at low WIMP masses, $m_X \lesssim 1$ GeV, with $\varepsilon  \lesssim 10^{-6}$, the parameter space is still largely open.

The situation becomes much more interesting once the WIMP is not millicharged but instead interacts with nucleons via a light mediator, in which case the CMB constraints are weakened. A well-studied realization of this is offered by photon kinetic mixing with a light vector mediator, $\varepsilon F'_{\mu \nu} F^{\mu \nu}$, where $F'_{\mu \nu}$ is the field strength of the light vector.  
As mentioned in Sec.~\ref{sec:nuc}, in that case the cross section between DM and nuclei will be given by Eq.~(\ref{m02}) where $\mu_0= \text{max}(m_{\phi}, a^{-1})$.  
 If the mass of the mediator is much larger that the temperature of the recombination era $\sim$ eV, by the 
time of recombination WIMP-nucleon interactions have become effectively of contact type and the CMB constraints are substantially weakened~\cite{Dvorkin:2013cea}.
Although the mediator is massive in this scenario, from the point of view of direct detection (because most experiments have a threshold energy larger than roughly 1 keV), mediators with masses up to MeV lead to scattering between WIMPs and nuclei that can effectively be treated as 
if they were mediated by a massless particle. We have estimated that the introduction of a mediator mass from the eV to the MeV scale reduces the nuclear stopping power only by a factor of a few. In particular, this reduction of the stopping power
 corresponds to an increase in the value of $\varepsilon$, by a factor less than 3 (with the precise value depending on the mediator mass). Therefore, for mediators between roughly eV to MeV, 
one can still use the top panel of Fig.~2 keeping in mind that the CMB constraints become significantly weaker. Notice that our $\varepsilon$ is related to the kinetic mixing parameter $\varepsilon$ simply by rescaling $\varepsilon \rightarrow \varepsilon \sqrt{\alpha_{X}/\alpha_{EM}}$, where $\alpha_{X}$ is the dark fine structure constant. The green dashed line shows the CMB constraint for $m_{\phi}=1$ MeV. The constraint becomes stronger with lower values of $m_{\phi}$, eventually reaching the millicharged case i.e. $m_{\phi}=0$ (blue dashed line). For massive mediators one should make sure that the extra massive particles do not increase significantly the relativistic degrees of freedom during the Big Bang Nucleosynthesis (BBN) era. This is achieved if we demand that the mediator decays before the universe was 3 seconds old. Following~\cite{Kaplinghat:2013yxa} we find the minimum value of $\varepsilon$ that leads to decays without affecting BBN. The solid green line shows that value for $m_{\phi}=1$ MeV. The constraint on $\varepsilon$ scales as $m_{\phi}^{-1/2}$. 

Additional constraints become relevant for the 1 MeV mediator case. Firstly, there is a limit on the kinetic mixing $\varepsilon \lesssim 4\times 10^{-4}$~\cite{Harnik:2012ni} (for $m_{\phi}$ between 0.5 and 1 MeV). Using that and requiring  that $\alpha_{X}$ satisfies the constraints on self-interactions from the Bullet Cluster~\cite{Randall:2007ph}, we obtain a constraint on our $\varepsilon$ parameter shown as a black curve in the top panel of Fig.~2. We have used the analytic formulae of~\cite{Tulin:2013teo} in order to derive the appropriate constraint on $\alpha_{X}$ from the bullet cluster.

\begin{figure}[t]
\begin{center}
\includegraphics[width=.45\textwidth, height=0.4 \textwidth]{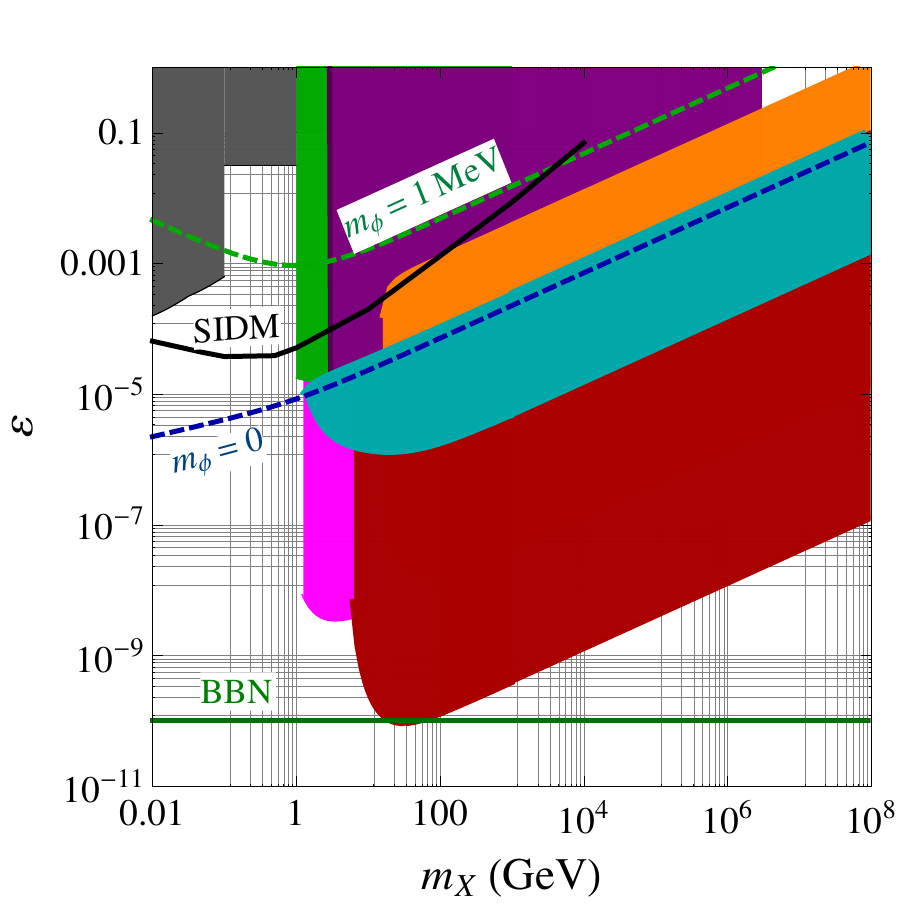}\quad \quad
\includegraphics[width=.45\textwidth, height=0.4\textwidth]{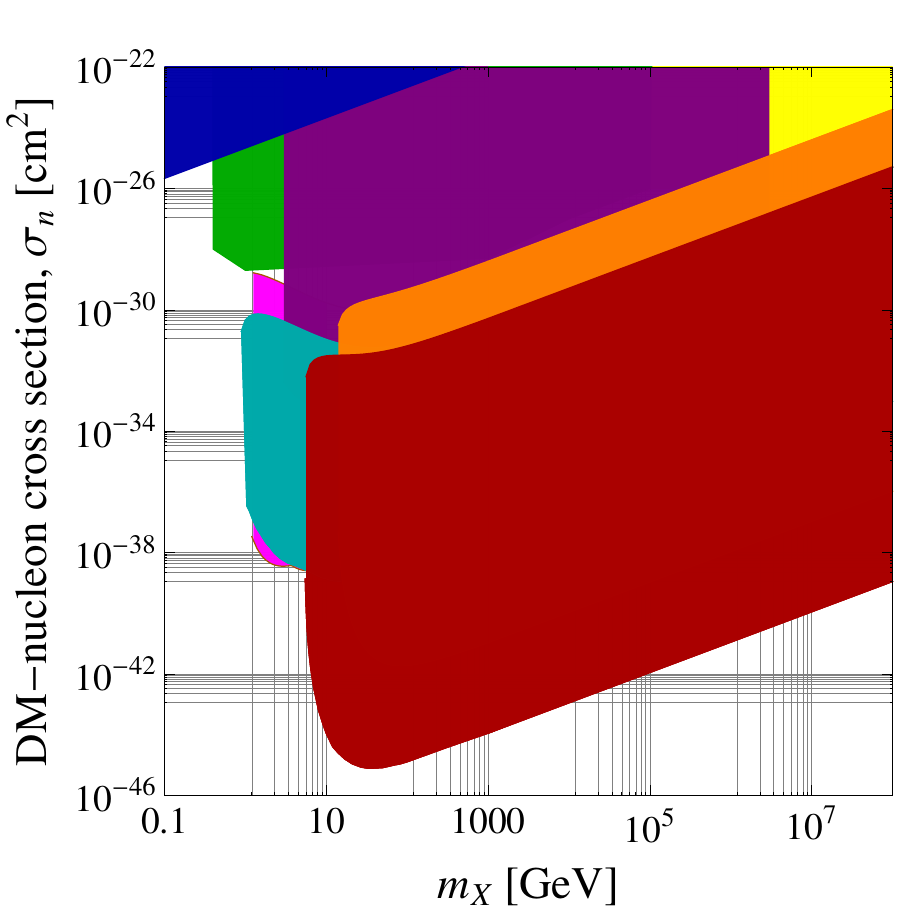}
\caption{The allowed $(\varepsilon-m_X)$ parameter space for millicharged and light mediators in the top panel, and contact interactions in the $(\sigma_{n}-m_X)$ plane in the bottom panel based on LUX (red), CRESST-1 (cyan), RRS balloon (purple), DAMIC~\cite{Barreto:2011zu} (magenta), XQC~\cite{Erickcek:2007jv} (green), CDMS-1 (orange), and the SLAC accelerator limits~\cite{Prinz:1998ua,Davidson:2000hf} on millicharged particles (gray). In the contact interaction case, we also include CMB limits~\cite{Dvorkin:2013cea} (blue), and those from the IMP8 experiment~\cite{IMP,Mack:2007xj} (yellow) constraints.}
\end{center}
\end{figure}

As one can see from the top panel of Fig.~2, the possibility of dark matter scattering off nuclei via exchange of light mediators  reveals a large new parameter space which becomes available and unconstrained even for relatively large values of the coupling $\varepsilon$ up to $10^{-3}$. It is also clear that even if LUX (or any other experiment at similar depth) lowers the energy threshold that triggers the detector, part of the parameter space will still be unconstrained simply because 
for strong enough WIMP-nucleus interactions, particles decelerate to energies that are always below the threshold. Notice also that experiments that are not susceptible to significant stopping (such as the balloon or rocket experiments) do no constrain the space below $m_X\sim 1$  GeV. The exploration of DM in this range of masses and interaction strengths can take place with a low-threshold detector in a much shallower site. Of course, the reason detectors have been planted in deep underground sites is because it is notoriously difficult to accurately model 
the backgrounds. Indeed, shallow site detectors will have an enhanced background of unwanted events. However, in the next section we briefly outline a proposal for exploiting large interaction strengths in order to carry out a search for a daily varying signal in shallow site detectors.

\begin{figure}[t]
\begin{center}
\includegraphics[width=.45\textwidth, height=0.4 \textwidth]{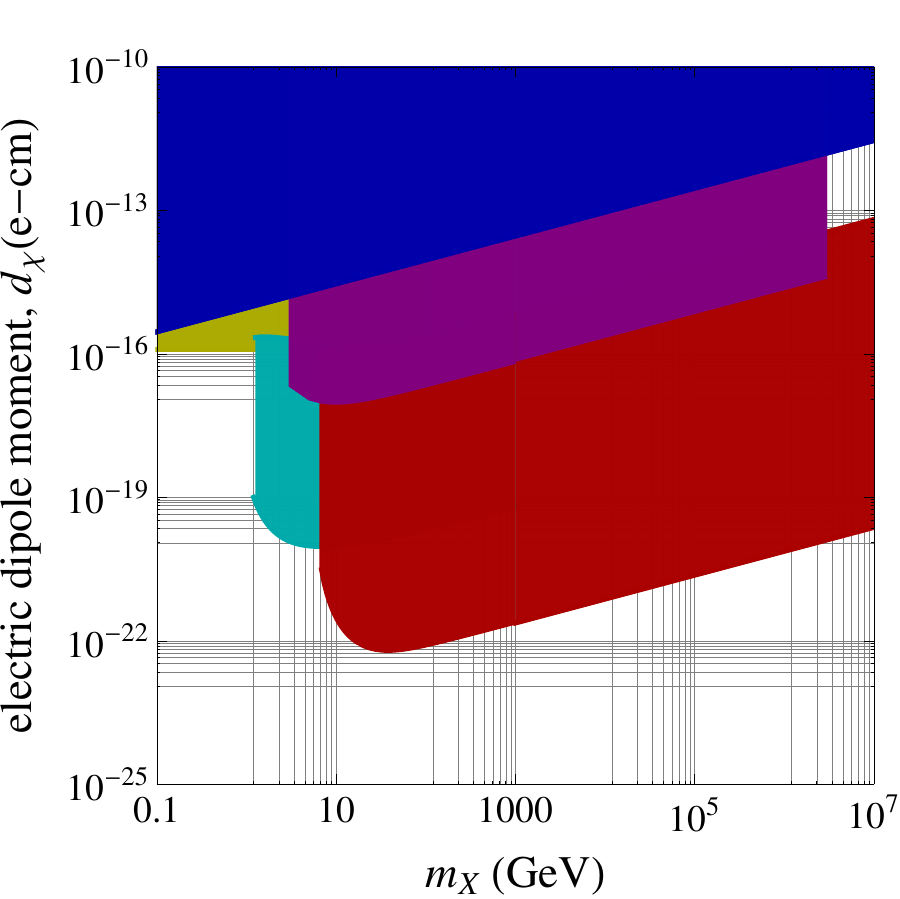}\quad \quad
\includegraphics[width=.45\textwidth, height=0.4\textwidth]{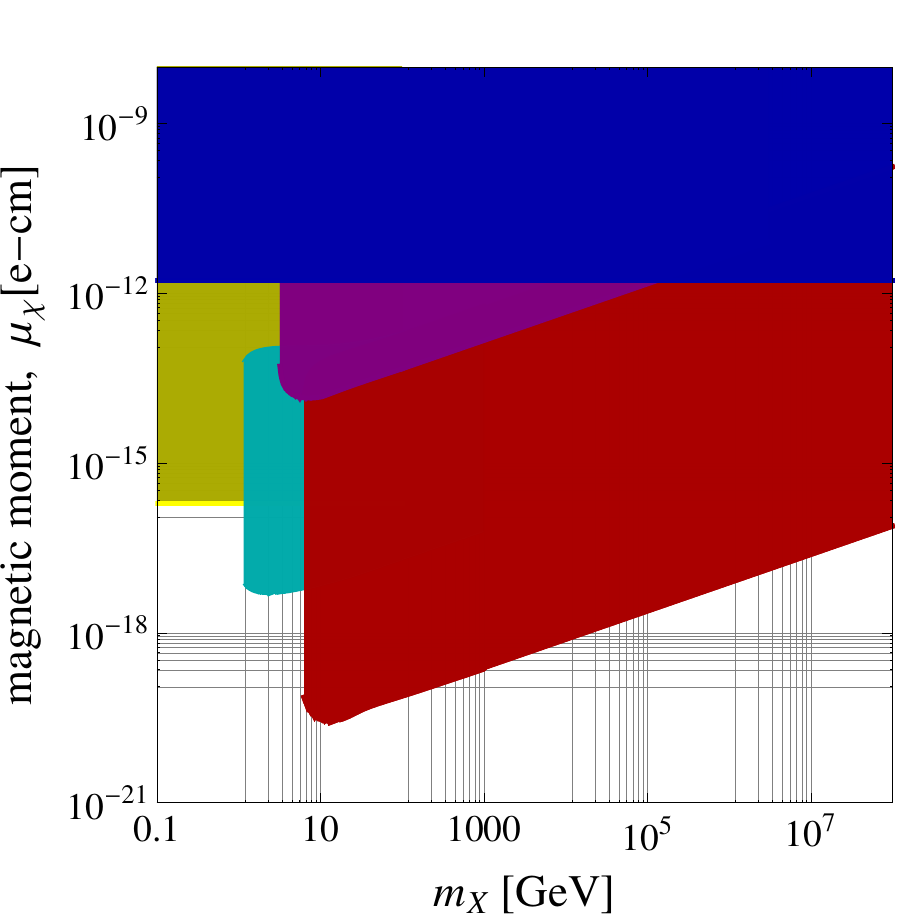} \label{dipoleplot}
\caption{The allowed/constrained electric dipole moment (top panel) and magnetic dipole moment (bottom panel) versus the dark matter mass. Here we have included the CRESST-1 (cyan), RRS balloon experiment (purple), LUX (red), CMB (dark blue), and LEP monophotons~\cite{Achard:2003tx,Fortin:2011hv} (dark yellow). }
\end{center}
\end{figure}

Let us turn now to contact interactions. In the bottom panel of Fig.~2 we plot the allowed parameter space for contact interactions i.e. WIMP-nucleon cross section versus $m_X$. One can see that although for large WIMP masses the intermediate regime between SIMPs and WIMPs has closed down since balloon and underground experiments have overlapping regions, for $m_X \lesssim 1$ GeV, this is not the case. As can be seen, XQC, CRESST and DAMIC are the main experiments constraining this region of the parameter space. Recall that CRESST data refer to the first much shallower site of CRESST ($\sim 1.4$ km). Given that reasonable doubts have been raised of how well the background of CRESST has been understood~\cite{Kuzniak:2012zm}, we urge the reader to interpret this region with care. Although a similar concern can be also raised for the CDMS shallow site data, CDMS does not exclude parameter space that has not been excluded by other experiments.  As in the case for light mediators, part of the parameter space is inaccessible to deep site detectors and therefore a different detection technique should be implemented. As before, daily modulated dark matter signal in shallow site detectors can be the way to probe these candidates.

In Fig.~3 we plot the allowed region for electric dipole (top panel) and magnetic dipole interactions (bottom panel) versus $m_X$. One can see that the large $m_X$ regime is excluded by different types of experiments unless one reduces significantly 
the WIMP-nucleus cross section (which corresponds to a reduction of the electric or the magnetic dipole moment). Deep sited detectors like LUX are perfectly suitable for constraining this part of the phase space since, for large enough dipole moments where
nuclear stopping is significant, the corresponding dark matter candidates are excluded mostly by balloon and CMB constraints that are not susceptible to stopping. On the other hand, for $m_X \lesssim 10$ GeV, the main experiments constraining the phase space \footnote{For significantly lighter dark matter particles from the range we have plotted a lot of other astrophysical, and cosmological constraints arise. For a detailed overview see e.g.~\cite{Davidson:2000hf}} are LEP, CMB, DAMIC, and CRESST. Note however that experiments with spin-dependent sensitivity, such as COUPP~\cite{Behnke:2012ys} and PICASSO~\cite{Archambault:2012pm}, are also relevant for light DM with magnetic dipole interactions. Additional complementary constraints are offered by helioseismology data  which is especially relevant for asymmetric DM models since very large DM abundances can accumulate in the Sun (see e.g.~\cite{Lopes:2013xua,Vincent:2013lua,Lopes:2014aoa}). 

 \section{Diurnal Modulation}
 
As we have argued in the previous section, dark matter particles with masses below 10 GeV that interact with nuclei via exchange of mediators with mass between eV to MeV,  or with contact interactions, can be unconstrained both by high altitude and underground experiments. Moreover, as we argued, even with a lower detection threshold, detectors deep underground will be unable to probe the entire allowed parameter space simply because the nuclear stopping power will have decelerated the dark matter particles
to insignificant kinetic energies that lie below threshold. Therefore this part of the parameter space can only be explored by shallow site detectors that have the disadvantage of having large unwanted backgrounds, thereby making dark matter detection a difficult task. However, this interesting part of the parameter space can be detected in principle if one looks for a daily varying modulated signal in shallow site detectors, arising simply from the fact that as the Earth rotates around its axis, dark matter particles 
travel different distances underground before reaching the detector. It is expected that if and when the detector aligns maximally with the Earth's velocity with respect to the dark matter halo,  because of the dark matter wind, there will be more particles trying to reach the detector from above. The shallow location insures this to be possible. However, one should expect that as the Earth rotates, there will be more dark matter particles trying to reach the detector not from above but in the worst case from below. However the flux of dark matter particles reaching from below will be hugely attenuated due to the stopping effect, thus creating a daily modulated signal. The possibility of diurnal modulated signals has been suggested earlier in the context of SIMPs \cite{Collar:1992qc,Hasenbalg:1997hs} and  of mirror dark matter~\cite{Foot:2011fh}.

Let us consider in some detail the daily variation of the dark matter signal in a detector located in a  shallow site. If $\hat{n}$ is the unit vector with direction from the center of the Earth to the detector, $v_E$ is the velocity of the Earth in the rest frame of the galaxy, $\theta_l$ is the latitude of the detector, and $\alpha$ the angle between $\vec{v}_E$ and  the angular velocity $\vec{\omega}$ of the Earth, in a galactic rest frame where the axis-$z$ is along the north-south pole, and $\vec{v}_E$ lies along the $x-z$ plane, we have the following relations
\be
\hat{n}=\hat{x} \cos \theta_l \cos\omega t+ \hat{y}\cos\theta_L \sin\omega t \pm \hat{z} \sin\theta_l, 
\ee
\be
\hat{v}_E=\hat{x} \sin \alpha+ \hat{z} \cos \alpha,
\ee
where the $\pm$ corresponds to the north and south hemisphere. We have chosen $t=0$ the time where $\vec{v}_E$ and $\hat{n}$ align as much as possible i.e. $\hat{n}$ is along the $x-z$ plane. Since the Earth is moving with respect to the rest frame of the galaxy, the Maxwell-Boltzmann distribution of dark matter is now shifted as
\be 
f(v)d^3v=\left (\frac{1}{\pi v_0^2} \right )^{3/2}e^{-\frac{v^2+v_E^2}{v_0^2}}e^{-\frac{2vv_E}{v_0^2}\cos\delta}d^3v \label{distri},
\ee
where $\delta$ is the angle between $\vec{v}$ and $\vec{v}_E$. In order to find $\delta$ we express the WIMP velocity $\vec{v}$ as
\be
\vec{v}=v(\hat{x} \sin\theta \cos\phi +\hat{y} \sin\theta \sin\phi + \hat{z}\cos\theta),
\ee
where we use the usual polar angles $\theta$ and $\phi$ to characterize $\vec{v}$. 
The angle $\delta$ now reads
\be
\cos\delta=\hat{v} \cdot \hat{v}_E=\sin\alpha \sin\theta \cos\phi +\cos\alpha \cos\theta. \label{delta}
\ee
Let us now define $\psi$ to be the angle between $\vec{v}$ and $\hat{n}$.
The angle $\psi$ is given in terms of known angles as
\begin{eqnarray}
\cos\psi=\hat{v}\cdot \hat{n}=\cos\theta_l \cos\omega t \sin \theta \cos\phi \nonumber\\
+ \cos\theta_l \sin\omega t \sin\theta \sin\phi \pm \sin\theta_l \cos\theta. \label{psii}
\end{eqnarray}
For a given WIMP velocity, the distance that has been traveled by the WIMP inside the Earth is 
\bea
 \ell(\psi) &=& \left( R_{\oplus}-\ell_{D} \right)\cos \psi  \label{length} \\ \nonumber 
&+& \sqrt{(R_{\oplus}-\ell_{D})^{2} \cos^{2} \psi - (\ell_{D}^{2} - 2 R_{\oplus}\ell_{D})} 
\eea
where $R_{\oplus}$ is the radius of the Earth, and $\ell_{D}$ the depth that the detector is located. If the distance required to stop the particles underground $\ell_{0}$ is smaller than the diameter of the Earth, the signal will be approximately proportional to a step function $\Theta(1- \frac{\ell(\psi)}{\ell_{0}})$.

The rate of dark matter events captured in a detector should be 
\bea
\frac{dR}{dE_R}&=&\\ \nonumber
&N_T&\frac{\rho_{X}}{m_X}\int_{v_{min}}^{v_{esc}}\frac{d\sigma}{dE}(E',E_R)v^3 f(v)dvd\cos\theta d\phi,
\eea
where $N_T$ is the number of scatterers in the detector, $f(v)$ is given by Eq.~(\ref{distri}), and $\frac{d\sigma}{dE}(E',E_R)$ is chosen accordingly among Eqs.~(\ref{m2}), (\ref{cross_electric}), and (\ref{cross_magnetic}) depending on the interaction case (millicharged, contact, electric, magnetic dipole) having substituted $E$ by $E'$. For a given velocity $v$ (and consequently $E$), $E'$ is given by $E_f$ from Eqs.~(\ref{m5}), (\ref{stop_electric}), or (\ref{stop_magnetic}) depending on the type of WIMP substituting $\ell (\psi)$ of Eq.~(\ref{length}) for $L$. The minimum velocity $v_{min}$ to create a recoil $E_R$ can be estimated as follows: just before the collision with the detector, the WIMP must have an energy at least $E_R/\gamma$. Using again the appropriate equation among (\ref{m5}), (\ref{stop_electric}), and (\ref{stop_magnetic}), with $E_f=E_R/\gamma$, and $L=\ell (\psi)$, one can find $E_{in}$ and correspondingly $v_{min}=\sqrt{2E_{in}/m_X}$.  We should emphasize here that although we have chosen to describe the rate in terms of the WIMP incident velocity before the particle goes underground and begins decelerating, the flux is correctly going to be proportional to $v$ even after deceleration. One can easily realize that although particles might decelerate, due to continuity of the number of particles crossing the surface of the Earth, the flux will not change and will be as if there was no deceleration underground. Note also that the daily modulation comes from the fact that as the Earth rotates around its own axis, the distance traveled underground by the WIMPs changes as a function of $\psi$ which depends on $t$ as it can be seen in Eq.~(\ref{psii}).

Due to the rotation of the Earth around the Sun, the angle $\alpha$ varies between the values $36.3^{\circ}$ (October 25th) to $49.3^{\circ}$ (April 25th) within the year. It is easy to demonstrate heuristically that the best site in order to maximize the diurnal modulation signal is roughly a location in the south hemisphere with
$\theta_l\simeq-\alpha$. At that  location when the detector aligns maximally with the WIMP wind, the attenuation due to stopping is at a minimum since the wind becomes tangential to the detector's location while twelve hours later, WIMPs from the wind have to travel several kilometers underground and therefore as long as WIMPs have decelerated sufficiently, the attenuation of the signal is maximum. It is important to note that in principle there are cases where $\theta=\pi/2-\alpha$ might be a better choice. This will occur if the distance required to effectively stop the particles is close to the diameter of the Earth. In that case a detector located at $\theta=\pi/2-\alpha$
maximizes the attenuation at $t=12$ hours (better than a detector at $\theta=-\alpha$) because it makes use of the full diameter of the Earth to stop the particles, while there is no attenuation at $t=0$ because the WIMPs in this case need several kilometers to decelerate effectively. However this latter case is not of interest for us. In this case, the vast majority of the WIMPs will arrive in underground detectors like LUX with no significant energy loss and therefore candidates like these are susceptible to the severe LUX constraints.  If we consider an annually average value of $\alpha=43^{\circ}$, the most appropriate location for such a detector will be in southern Argentina (for example Sierra Grande as suggested previously~\cite{Hasenbalg:1997hs,Foot:2011fh}), Chile, or New Zealand. We find for example that at such a location, a 1 GeV DM mass with a contact cross section $\sigma_{p} = 10^{-30}~{\rm cm}^{2}$ (allowed by Fig. 1) exhibits a $\sim 79 \%$ daily modulation in the total rate on a silicon target with 0.4 keV threshold.

\section{Conclusions}

In this paper we have studied in detail the deceleration of dark matter particles  due to interactions with terrestrial atoms and what impact this has on direct dark matter detection. In particular, we have studied the case of dark matter-atom interactions of long-range (massless or low mass mediators), of contact and of electric and magnetic dipole types. We find that in the case where the dark matter particle is light (less than 1 GeV) and the interactions is either contact or mediated by light (but not massless) particles, there is parameter phase space that cannot be probed by current underground detectors even with substantially lowered energy thresholds. This region of the parameter space can be probed by shallow site detectors with low energy thresholds.  However, since in this case dark matter particles will be very effectively stopped if coming upwards (i.e.  below the detector), we argue that a search for a daily modulated dark matter signal is probably the best strategy for probing this part of the parameter space. An alternative will be a detector that will be able to distinguish the direction of the scattered dark matter particle (see e.g. DRIFT~\cite{Morgan:2003qp}, DMTPC~\cite{Dujmic:2007bd}, MIMAC~\cite{Santos:2007ga}). 

It is important to stress as well open directions for future work. We have throughout this work assumed that DM-nuclear scattering proceeds elastically. A detailed study of nuclear stopping for inelastically scattering DM is beyond the scope of this work, but may be interesting to return to in the future. We note that a diurnal search for magnetically interacting inelastic DM was recently proposed~\cite{Pospelov:2013nea}. In addition to the elastic scattering assumption, we have also assumed (for contact and light mediator interactions) isospin-conserving interactions, though this need not be the case~\cite{Kurylov:2003ra,Giuliani:2005my}. The effects of nuclear stopping could for example be significantly reduced for ''oxygen-phobic'' interactions ($f_{n}/f_{p} \simeq -1$). 

Lastly, the diurnal signal proposed here at low DM masses requires nuclear recoil energy thresholds $\lesssim $ keV. DAMIC has already demonstrated this capability and can likely derive limits at lower DM masses as the Si quenching is better understood~\footnote{We note however that the CCD technology used by DAMIC is unlikely to be useful for a diurnal search given the lack of timing information.}. In addition, DM-electron scattering allows DM masses down to $\sim$ MeV to be probed~\cite{Essig:2011nj}. Thus for example the photon kinetic mixing model (see the upper panel of Fig.~2) can be further probed in such searches. It is important to stress however that the existing search for $e^{-}-{\rm DM}$ scattering with XENON10 data~\cite{Essig:2012yx} is not as sensitive to the diurnal modulation signal we propose here as a surface detector. We hope to return to this in future work.
\\
\\

\begin{centering}
{\bf ACKNOWLEDGEMENTS \\}
\end{centering}
\vspace{.4cm}
We are more than grateful to Peter Sigmund for countless discussions and his patience in explaining the mechanisms of stopping power to us. We would also like to thank E. Lidorikis for clarifying our questions regarding graphite. The CP3-Origins centre is partially funded by the Danish National Research Foundation, grant number DNRF90.

\vspace{1cm}

\newpage

{\bf \hspace{3cm} APPENDIX}

 \subsection{RRS Ballon Flight}
Rich, Rocchia and Spiro (RRS)~\cite{Rich:1987st} derived DM-nucleon limits from a short balloon flight with a low-threshold silicon detector in 1977. Though the detector was 0.5 g in mass, the exposure time was not specified. It can however be reasonably well inferred from the spectrum in the energy interval $E_{R} =$ [0.5-7] keV presented in their Fig. 1. Under the assumption that their highest energy bin contains $\sim$1 event we can normalize their spectrum and infer an approximate exposure of 16.7 g-days. We then derive limits on the DM-nucleon cross section in the various models considered in the text by requiring that a given parameter point not produce more than the total event that they reported in their Fig. 1. We find that this procedure reproduces the limits reported in RRS under the contact interaction assumption. The effects of the atmosphere's stopping is not included here, but can be derived from the 4.5 $g~{\rm cm}^{-2}$ of column density at the maximum altitude of the balloon. 

\subsection{CRESST-I}
The first CRESST run was obtained from a 1.51 kg-day exposure on a sapphire target, Al$_{2}$O$_{3}$~\cite{Altmann:2001ax}. Located at the Gran Sasso Underground Laboratory the experiment was located 1.4 km underground. The experiment operated with the very low energy threshold of $600$ eV. We have found that the effective efficiency is well approximated by a flat 30 $\%$. 

\subsection{CDMS-I}
The first stage of the CDMS experiment operated at the Stanford Underground Facility in a tunnel 10.6 m underground~\cite{Abrams:2002nb}. The collaboration obtained a 15.8 kg-day exposure on a combination of Si and Ge targets with a rather high energy threshold of 25 keV. We include the 27 and 4 events from neutron backgrounds on Ge and Si respectively to obtain our limits with an effective efficiency of 50 $\%$.

\subsection{LUX}
To obtain the LUX limits we follow the method used in~\cite{Cirigliano:2013zta,Frandsen:2014ima} which well approximates LUX's low-energy sensitivity which is crucial to obtaining reliable exclusion limits at low masses.  For completeness we briefly review the method here. The first LUX run~\cite{Akerib:2013tjd} had an exposure of 10,065 kg-days. The collaboration quotes an upper limit of 2.4 signal events for $m_X <$ 10 GeV which we conservatively apply to the whole mass range though stronger limits can be obtained. Following the collaboration we employ a sharp cutoff at 3 keV in the light yield so as to remove uncertainties from the analysis that become sizeable at lower energies.

 \subsection{DAMIC}
 
 The DAMIC detector had a 107 g-day exposure on a silicon target with a very low 40 ${\rm eV}_{ee}$ energy threshold at Fermilab's NuMI site, roughly $\sim 107$ m underground. We follow the DAMIC collaboration by using the Lindhard theory to compute the quenching factor at low energies. We have found that a flat 15$\%$ detector efficiency well approximates their set of selection cuts. 
 
\subsection{LEP}
The LEP experiment~\cite{Achard:2003tx} has reported limits on mono-photon plus missing energy events which can be interpreted as limits on DM dipoles~\cite{Fortin:2011hv}. These limits are stronger than those expected from the 5$\sigma$ reach of a 14 TeV LHC. 

\subsection{BBN} 
To avoid strong BBN constraints on the massive dark photon model, we require $\varepsilon \gtrsim 10^{-10}$ (for a 1 MeV mediator), such that this mediator has decayed away within the first 3 seconds of the Universe's history~\cite{Kaplinghat:2013yxa}.

\subsection{CMB}
In the case of a 1 MeV mediator, we can repurpose the limits of~\cite{Dvorkin:2013cea} obtained under the assumption of a contact interaction, which should be accurate at the recombination time. The CMB limits on millicharged DM are taken from Eq. (3) of ~\cite{Dolgov:2013una}.

\end{document}